\documentclass[useAMS,usenatbib]{mn2e}
\usepackage{graphicx}
\def\beg{\begin{eqnarray}}
\def\ende{\end{eqnarray}}
\def\gsim{\lower.4ex\hbox{$\;\buildrel >\over{\scriptstyle\sim}\;$}} 
\def\lsim{\lower.4ex\hbox{$\;\buildrel <\over{\scriptstyle\sim}\;$}}

\newcommand{\Pm}{\mbox{Pm}}
\newcommand{\Rm}{\mbox{Rm}}
\renewcommand{\vec}[1]{\mbox{\boldmath $#1$}}
 
\def\Om{{\it \Omega}}
\def\R{{R\"udiger}}

\title[The magnetorotational instability of toroidal fields ]
{Astrophysical and experimental  implications from the magnetorotational instability of toroidal fields}
\author[G. R\"udiger, M. Gellert, M. Schultz, R. Hollerbach and F. Stefani]
{G. R\"udiger$^{1,4}$, M. Gellert$^1$, M. Schultz$^1$, R. Hollerbach$^{2,3}$, F. Stefani$^4$\\
$^1$Leibniz-Institut f\"ur Astrophysik Potsdam, An der Sternwarte 16, 14482 Potsdam, Germany\\
$^2$ETH Z\"urich, Institut f\"ur Geophysik, Sonneggstr 5, 8092 Z\"urich, Switzerland\\
$^3$University of Leeds, School of Mathematics, Woodhouse Lane, Leeds, LS2 9JT, UK\\
$^4$Helmholtz-Zentrum Dresden-Rossendorf, POB 51 01 19, 01314 Dresden, Germany}
\begin{document}

\date{Accepted . Received ; in original form }

\pagerange{\pageref{firstpage}--\pageref{lastpage}} \pubyear{2009}

\maketitle

\label{firstpage}

\begin{abstract}
The interaction of differential rotation and toroidal fields that are
current-free in the gap between two corotating axially unbounded cylinders is considered. 
It is shown that nonaxisymmetric perturbations are unstable if the rotation rate
and Alfv\'en frequency of the field are of the same order, almost  independent of the
magnetic Prandtl number $\rm Pm$.
For the very steep rotation law $\Om\propto R^{-2}$ (the Rayleigh limit) and for small $\rm Pm$ the threshold values of rotation and field for  this
Azimuthal MagnetoRotational Instability (AMRI)  scale with the ordinary Reynolds
number and the Hartmann number, resp.  A laboratory experiment {with liquid
metals like sodium or gallium in a Taylor-Couette container has been designed on the basis of this finding. For fluids with more flat rotation laws  the Reynolds number and the Hartmann number are no longer typical quantities for  the instability}.

For the weakly nonlinear system the numerical values of the kinetic energy and the
magnetic energy are derived for magnetic Prandtl numbers $\leq 1$.
We find that the magnetic energy grows monotonically
 with the magnetic Reynolds number $\rm Rm$,
while the kinetic energy   grows with $\rm Rm/\sqrt{\rm Pm}$.
The resulting turbulent Schmidt number, as the ratio of the `eddy' viscosity and the diffusion coefficient of a passive scalar (such as lithium) is of order 20 for
$\rm Pm=1$, but for small $\rm Pm$ it drops to order unity.
Hence, in a stellar core with fossil fields and steep rotation law the transport of
angular momentum by AMRI is always accompanied by an intense mixing of the plasma,
until the rotation becomes rigid.
\end{abstract}


\section{Introduction}

A hydrodynamically stable rotation law may become unstable under the presence of a
sufficiently strong uniform axial field.
For this so-called magnetorotational instability (MRI) the axial field provides the
catalyst which makes the rotation law unstable without any modification of the
uniform field (Velikhov 1959).
Of course, by this mechanism a protoplanetary disk with a Keplerian rotation law
should also become unstable if it is not too cold -- or (almost) equivalently -- if
the field is strong enough.
For an estimation of the critical magnetic field strength it is enough to use the
condition ${\rm S}\gsim 1$, where the Lundquist number
$
{\rm S}={B_z \, H}/{\sqrt{\mu_0\rho} \eta},
$
with $B_z$ the amplitude of the axial field, $H$ the semi-thickness of the disk,
and $\eta$ the microscopic magnetic diffusivity.
Plotted in the ${\rm S}$-${\rm Rm}$ plane, with
$
{\rm Rm}= {\Om \, H^2}/{\eta}
$
as the magnetic Reynolds number, and for small magnetic Prandtl number
\begin{equation}
{\rm Pm}=\nu/\eta,
\label{Pm}
\end{equation} 
the viscosity hardly influences the instability map.
At a distance of 1 AU from a central mass with 1 ${\rm M}_\odot$ the disk
thickness is about 3\%, the density $\rho$ of the gas $10^{-10}$ g/cm$^3$, and
$\eta \simeq 4\cdot 10^{15}$ cm$^2$/s (because of the low temperature of the gas).
Hence, the surprisingly high value of 0.1 G is obtained for the minimum $B_z$.

In order to excite nonaxisymmetric MRI modes one needs even higher magnetic fields.
Kitchatinov \& R\"udiger (2010) showed in a linear theory that nonaxisymmetric modes
only arise if $\Om_{\rm A}\gsim 0.05\, \Om$, with the Alfv\'en frequency
$\Om_{\rm A}=B_z/\sqrt{\mu_0 \, \rho} \, H$.
For $\Om=2\cdot 10^{-7}$ at 1 AU this condition is fulfilled for $B_z\gsim 0.3$ G.
Such strong fields at this distance cannot be due to a stellar dipole field in the center.
On the other hand, one should mention that the fossil magnetic fields found in meteorites
are of just this order.
One solution of this dilemma is to consider toroidal fields, which by the stretching
action of differential rotation may often exceed the values of the poloidal field.
For protoplanetary disks, however, the magnetic Reynolds number ${\rm Rm}$ only reaches
values of about 50, so that the resulting toroidal field does {\em not} dominate the
original $B_z$.

The same is true for galaxies when the interstellar turbulence is driven by SN-explosions.
The resulting magnetic Reynolds number is then also of order 10--100, so that the induced toroidal field does not dominate the poloidal one -- as observed.

A very different situation holds for the radiative cores of rotating stars.
Due to the high conductivity of the hot plasma the magnetic Reynolds number is so large
that even very slight differential rotation will produce toroidal fields which clearly
dominate the fossil poloidal fields.
It should then be sufficient to consider the stability of only the strong toroidal field
under the presence of differential rotation.
In the present paper we shall always apply a rotation law close to the profile of
uniform specific angular momentum, i.e. $\Om \propto R^{-2}$.
This rather steep profile is just beyond the regime of hydrodynamic centrifugal
instability, also in spherical systems (see R\"udiger \& Kitchatinov 1996).

There are basic stability conditions in cylindrical geometry.
A flow is stable against axisymmetric perturbations if
\begin{equation}
\frac{{\rm d}}{{\rm d}R}\left(R^2 \Om\right)^2 >
 \frac{R^4}{\mu_0\rho}\frac{{\rm d}}{{\rm d}R} \left(\frac{B_\phi^2}{R^2}\right)
\label{flow}
\end{equation}
(Michael 1954).
This condition turns into the well-known Rayleigh criterion for both
$B_\phi=0$ and $B_\phi\propto R$.
The latter case means that an axial uniform electric current does not suppress
Taylor vortices, which are due to too steep rotation profiles.
Hence, rotation laws which are steeper than $\Om\propto R^{-2}$ will always
excite axisymmetric rolls, even under the presence of very strong toroidal
fields with $B_\phi \propto R$.
On the other hand, any unstable $\Om$-profiles can be stabilized with a
toroidal field of sufficiently large amplitude and suitable geometry.

If nonaxisymmetric modes are considered then the necessary and sufficient
condition for stability becomes
\begin{equation}
\frac{{\rm d}}{{\rm d}R}\left(R\, B_\phi^2\right) < 0
\label{stab}
\end{equation}
(Tayler 1957, 1973, Vandakurov 1972), at least in the absence of rotation.
The fully general formulation including rotation is not known.
Two questions immediately arise: i) is the field corresponding to a uniform
current ($B_\phi \propto R$) really unstable, and ii) does the current-free
field $B_\phi \sim R^{-1}$ become unstable under the presence of
(differential) rotation? Both questions can be answered.
In the first case the resulting instability is called the Tayler instability
(TI), and in the second case, where fields and rotation profiles that would
each be stable on their own interact to yield instability, the result is
called the Azimuthal MagnetoRotational Instability (AMRI).
Whereas the energy for the TI comes from the electric current, the energy
source for AMRI is entirely from the differential rotation, and AMRI can only
saturate at the expense of the rotational shear by producing high values of
`eddy' viscosity.
This could therefore justify the high values of `artificial' viscosity which
Eggenberger, Montalb\'an \& Miglio (2012) introduced to explain the slow
rotation of the  stellar cores  after the collapse towards their red giant stage.

{The interaction of flow and field has already been considered by Chandrasekhar (1956) who proved that for ideal media the system $U_\phi(R) $ and $B_\phi(R)$ is {\em stable} for  $U_\phi(R) =B_\phi(R)$. Indeed, the current-free (AMRI) field $B_\phi\propto R^{-1}$ subject to the rotation with  $\Omega\propto R^{-2} $ (Rayleigh limit) belongs to this class. Instability is thus   only possible for $\Om\neq\Om_{\rm A}$ with the Alfv\'en frequency $\Om_{\rm A}=B_\phi/\sqrt{\mu_0 \rho R^2}$. Figure \ref{f0} (left) as a result of numerical calculations indeed shows the realization  of this finding for finite diffusivities. The lower  branches for $\rm Pm=1$ and $\rm Pm=10$ are very close to that limit. For very small $\rm Pm$ it is the upper branch which approaches the Chandrasekhar line $\Om=\Om_{\rm A}$ from below.}
\section{The  dispersion relation for $ \Om\propto R^{-2}$}
It is possible to demonstrate the mechanism which leads to AMRI with a simple
dispersion relation for the stability of {\em nonaxisymmetric} perturbations
in the form $\exp(\omega t +{\rm i}(k_z z + +k_R R + m\phi))$ with the real part $\Re(\omega)$ as the growth
rate. It is enough to apply the case with $\rm Pm\to 0$ and $\Om\propto 1/R^2$
(the Rayleigh limit) together with $B_\phi\propto R^{-1}$ to the expressions given by Kirillov, Stefani \& Fukumoto
(2012) for nonaxisymmetric perturbations. The resulting   dispersion
relation is of second order, i.e.
\begin{equation}
{\rm Re}^2 \hat\omega^2 + {\rm Re}\ (a_3
  + {\rm i}b_3)\hat\omega +a_4+ {\rm i}b_4=0.
\label{disp1}
\end{equation}
Here we have used the quantity $\hat\omega={\omega}/\Om$ as the growth rate in units of the angular velocity of rotation and the Reynolds number ${\rm Re}=\Om/\nu k^2$.
The imaginary part of  $\hat\omega$ describes an azimuthal drift $\hat\omega_{\rm dr}$ of the
nonaxisymmetric pattern (see below).
The coefficients in (\ref{disp1}) are
\begin{eqnarray}
&& a_3= 2(1+(2+m^{*2}{\rm Ha}^{*2})),
 \quad b_3=2m^* \, {\rm Re}^*\nonumber\\
&& a_4=1+(4+2m^{*2}) \, {\rm Ha}^{*2}+m^{*4} \,
 {\rm Ha}^{*4}-m^{*2} \, {\rm Re}^{*2}\nonumber\\
&& b_4=2{\rm Re}^* \, m^* (1+m^{*2} \, {\rm Ha}^{*2})
\label{disp2}
\end{eqnarray}
with the abbreviations 
\begin{equation}
{\rm Re}^* = \alpha \, {\rm Re}, \quad {\rm Ha}^*=
 \frac{\alpha}{k R}\, {\rm Ha}, \quad m^*=\frac{m}{\alpha},
\label{RE1}
\end{equation}
where $\alpha= k_z/k$ and ${\rm Ha}=B_\phi /(k\sqrt{\mu_0\rho\nu\eta})$.

The formal solution of (\ref{disp1}) for marginal instability leads to
the two conditions
\begin{equation}
\hat\omega_{\rm dr}= -\frac{b_4}{{\rm Re}\ a_3},
\quad\quad\ \ \ \ \ \ \ \ \ a_3(a_3 a_4 +b_3 b_4)=b_4^2.
\label{disp3}
\end{equation}
The first of these relations gives the negative drift rate in units of the
basic rotation, i.e.
\begin{equation}
\frac{\omega_{\rm dr}}{\Om}=
 - \frac{m (1+m^{*2}{\rm Ha}^{*2})}{1+(2+m^{*2}){\rm Ha}^{*2}}.
\label{disp4}
\end{equation}
The observable drift in the laboratory system, ${\rm d}\phi/{\rm d} t=
-\omega_{\rm dr}/m$, is thus positive (in the rotation direction) and with
${\rm d}\phi/{\rm d} t/\Om\leq 1$. {This value is reduced by medium Hartmann numbers but it is 
independent of the magnetic field for large Hartmann numbers.  Note the absence of  the magnetic Prandtl number in the relation (\ref{disp4}). For small $\rm Pm$ the drift values (and also the axial wave numbers) do not depend on the actual values of $\rm Pm$ (see Fig. \ref{f0}). The phase speed ${\rm d}\phi/{\rm d} t$ is also independent of  the sign of the mode number $m$.} 

The second condition in (\ref{disp3}) leads to 
\begin{equation}
{\rm Re}^*  \simeq \frac{m^{*3}{\rm Ha}^{*2}}{2},
\label{disp5}
\end{equation}
the latter relation holding for strong fields.  Axisymmetric solutions do not exist. It is
also clear  that the mode with $m=1$ is most easily excited so that
\begin{equation}
\frac{{\rm Ha}^2}{{\rm Re}}\simeq 2 {k_z^2 R^2},
\label{disp6}
\end{equation}
again valid for large $\rm Ha$. {The  influence of the axial wave number on the excitation condition is rather strong}. Solutions with large axial wave number --
representing oblate cells -- require much slower rotation to be excited than round cells.
For the Elsasser number ${\rm \Lambda}=B_\phi^2/\mu_0\rho \eta\Om$ one obtains
$
{\rm \Lambda} \simeq 2 (k_zR)^2,
$
{where the influence of both the microscopic viscosity and the radial wave number  completely disappear. Note that the normalized wave number $k_z R$ is of order unity for spheres or axially unbounded cylinders but is much larger for thin disks. Another interesting reformulation of this relation  is
\begin{equation}
\frac{\Om_{\rm A}^2}{\Om^2}\simeq \frac{2 k_z^2 R^2}{\rm Rm}
\label{disp7}
\end{equation}
with $\rm Rm=Pm\ Re$. It is thus clear that for large enough $\eta$, i.e. for sufficiently low electric conductivity, the ratio (\ref{disp7}) may exceed unity, which is not the case for $\rm Pm=1$ (see Fig. \ref{f0}, left). Hence, for small magnetic Prandtl number AMRI exists in the strong-field  limit ($\Om_{\rm A}>\Om$) while the opposite is true ($\Om_{\rm A}<\Om$) for $\rm Pm \geq 1$. Equation (\ref{disp7}) with $k_z\simeq \pi/H$ and with the above given disk parameters leads to a critical magnetic field of $B_\phi\simeq 3 R/H\simeq 100 $\ G, the latter for a thin disk with $H/R\simeq 0.03$. One needs about $10^{-2}$\ G poloidal field to induce  toroidal fields with 100\ G in such disks. 

For the inner radiative zones of hot giants  the relation (\ref{disp7}) turns into $\Omega_A^2\simeq 10 \eta \Omega$ which with $\rho\simeq 10 $ g/cm$^3$,  $\eta\simeq 10^2$ cm$^2$/s and $\Omega\simeq 10^{-5}$ s$^{-1}$ provides about 1 G as the threshold value of the  toroidal field.} 
\section{A hydromagnetic Taylor-Couette problem}
\begin{figure*}
 \includegraphics[width=5.7cm]{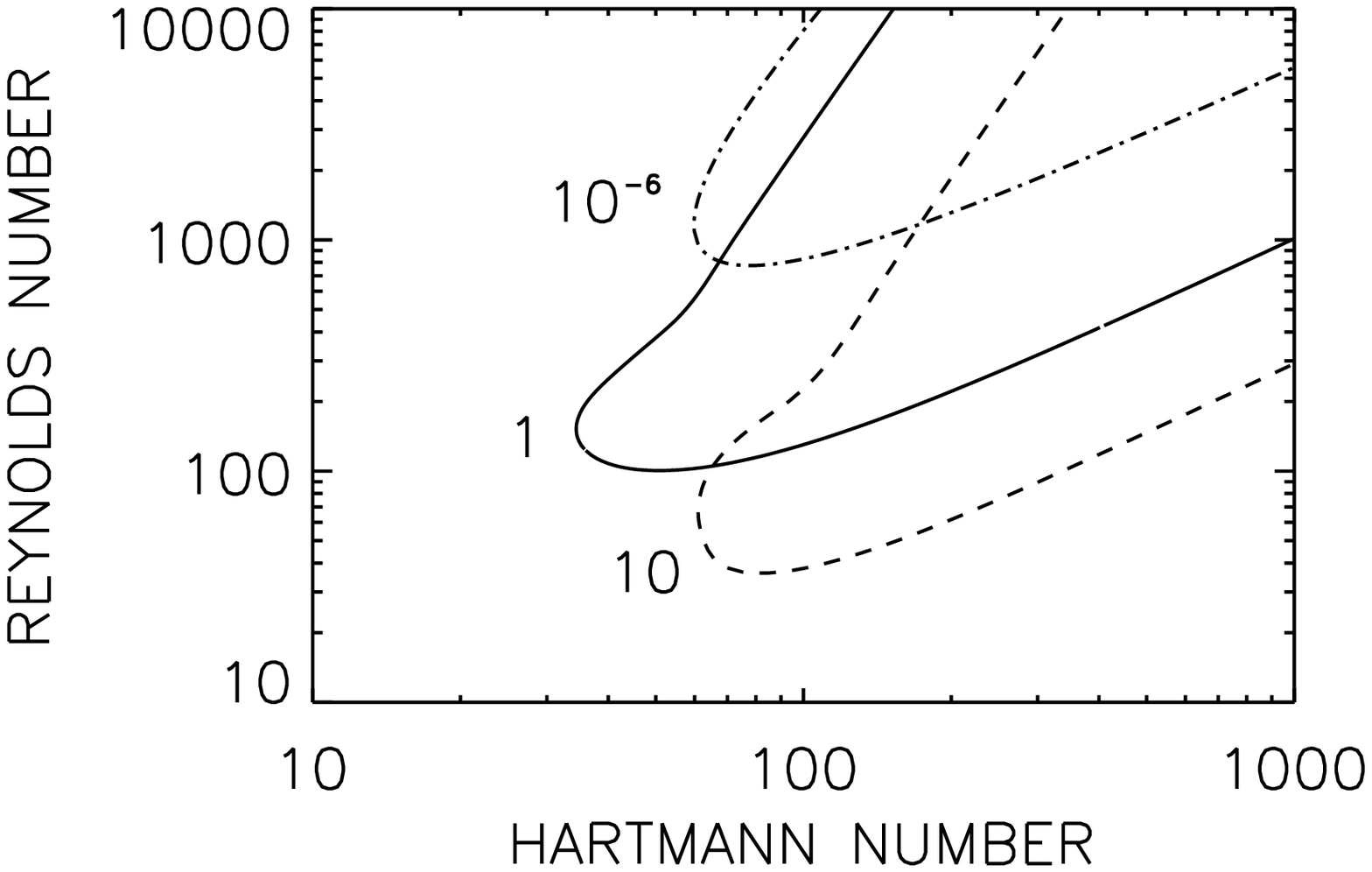}
 \includegraphics[width=5.7cm]{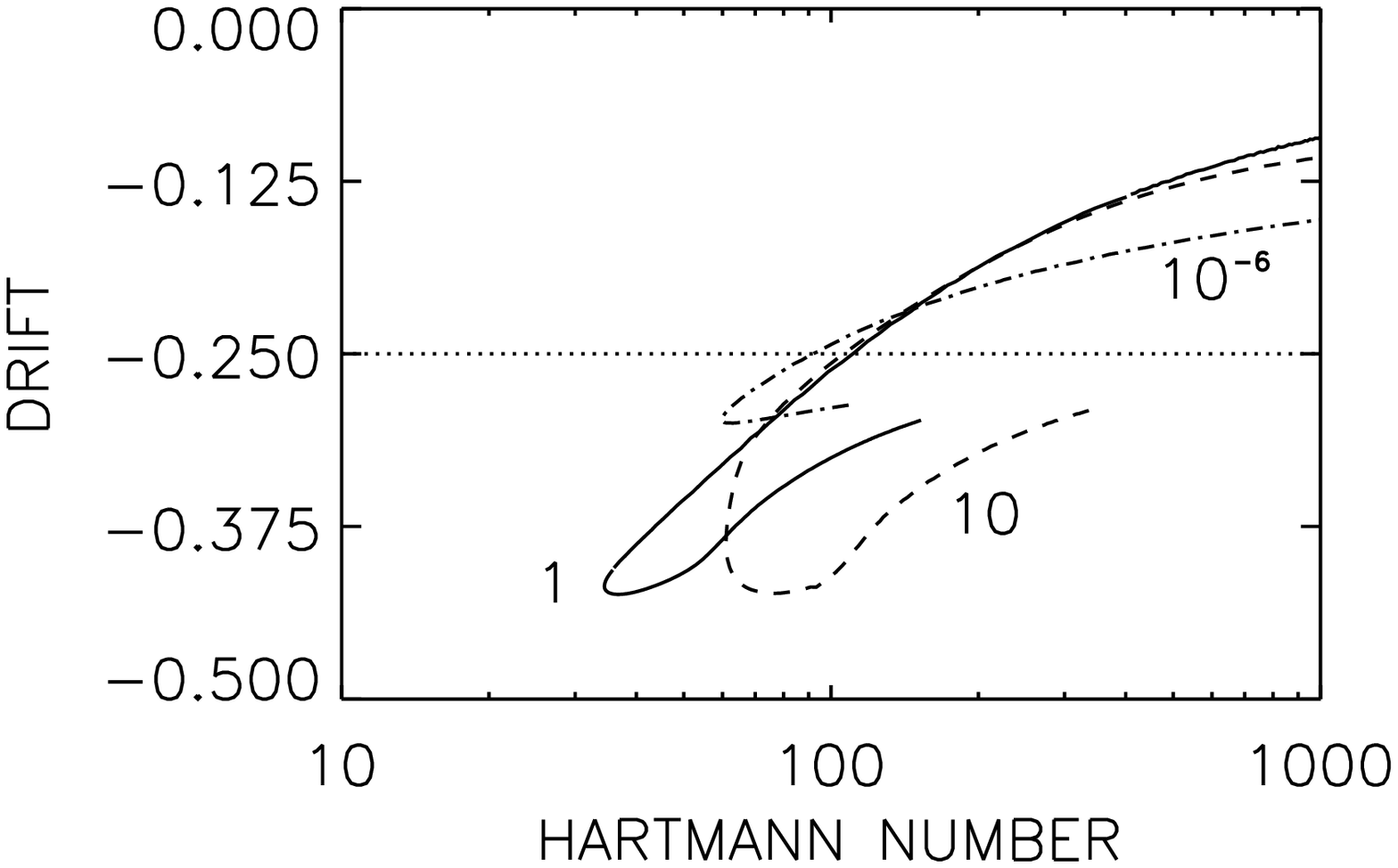}
 \includegraphics[width=5.7cm]{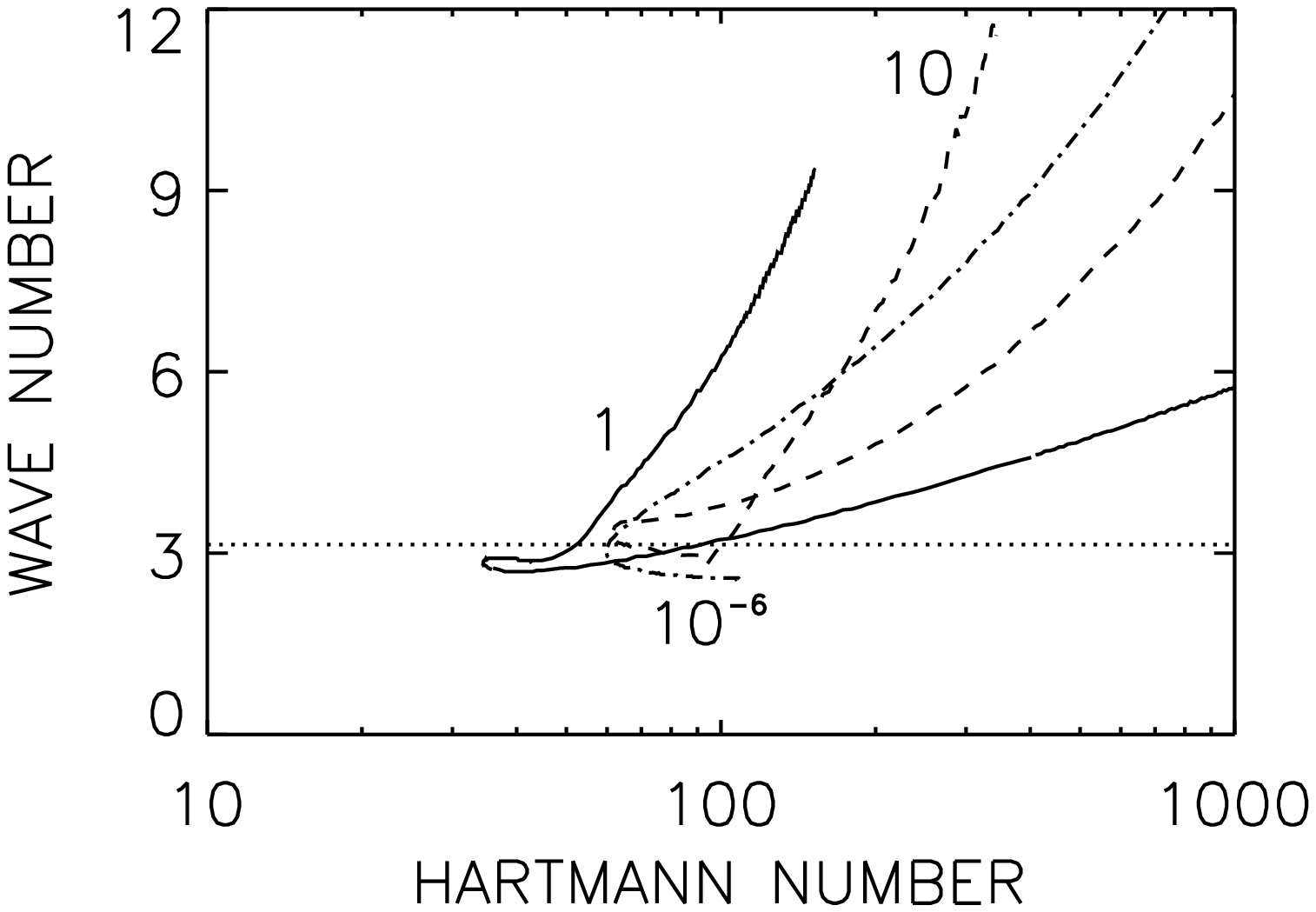}
 \caption{The AMRI at the Rayleigh limit ($R^2\Om={\rm const}$, i.e. $\mu_\Om=0.25$) for the  magnetic
Prandtl numbers $\rm Pm= 10^{-6}$, $\rm Pm=1$ and $\rm Pm=10$. Left: the instability map, 
middle: the azimuthal drift of the solutions normalized with $\Om_{\rm in}$,
right: the axial wave number $k$  multiplied with the gap width $D$. The Reynolds  numbers, the drift values and the wave numbers  for $\rm Pm< 10^{-4}$ cannot be distinguished from the  values at the given lines for $\rm Pm=10^{-6}$.}
\label{f0}
\end{figure*}

With view on the experimental demonstration of AMRI we consider
a Taylor-Couette set-up with two
corotating cylinders of nearly perfectly conducting material. The cylinders
confine an incompressible, conducting fluid under the presence of a toroidal
magnetic field. It is clear that the radial profiles of the rotating flow and
the field between the cylinders are
\begin{equation}
U_\phi= R \, \Om = a_\Om \, R+ \frac{b_\Om}{R},
 \quad\quad B_\phi=a_B R + \frac{b_B}{R}.
\label{uphi}
\end{equation}
Let the ratio of the rotation rates of the cylinders be
$\mu_\Om=\Om_{\rm out}/\Om_{\rm in}$ and a similar expression
for the field amplitudes, i.e.
\begin{equation}
\mu_B= \frac{B_{\rm out}}{B_{\rm in}}.
\label{mueB}
\end{equation}
The ratio of the cylinder radii is $r_{\rm in}=R_{\rm in}/R_{\rm out}$,
where we will fix $r_{\rm in}=0.5$. For such a container $\mu_\Om=0.25$
describes the Rayleigh limit for which $R^2\!\Om \simeq {\rm const}$
($a_\Om=0$ in Eq.~(\ref{uphi})), while $\mu_\Om=0.5$
describes the quasi-galactic rotation law $U_\phi\simeq {\rm const}$.
{These two rotation laws form the extremes which will be considered here in detail. As they behave rather differently the solutions for the rotation laws between them (e.g. the Kepler rotation law)   should be   more complex and might even depend on the boundary conditions. As a special choice for a laboratory experiment also a rotation law with $\mu_\Om=0.26$ has been considered which  is hydrodynamically stable but  close enough to the Rayleigh limit to become unstable with rather weak magnetic fields and slow basic rotation (see below).}

The magnetic  profile in (\ref{uphi}) also contains two extrema.
With $b_B=0$ the toroidal field is due to a homogeneous current inside the fluid.
For $r_{\rm in}=0.5$ the corresponding value is $\mu_B=2$. Such fields are
unstable according to the criterion (\ref{stab}) against nonaxisymmetric
perturbations. The existence of this kink-type `Tayler' instability has recently
been shown in the laboratory (Seilmayer et al. 2012, R\"udiger et al. 2012).

The equations and boundary conditions are given by \R\ et al.\ (2007) and will
not be repeated here. The cylinders, which are unbounded in the axial
direction, are imagined  to be made of a perfectly conducting material. The AMRI in the 
container with $r_{\rm in}=0.5$ appears for $\mu_B=0.5$. The electric current only flows within the
inner cylinder. 

{The main parameters of the theory are the Reynolds number and the Hartmann number defined by }
\begin{equation}
{\rm Re}= \frac{\Om_{\rm in}\, D ^2}{\nu},\ \ \ \ \ \ \ \ \ \ \ \ \ 
{\rm Ha}=\frac{B_{\rm in}\, D}{\sqrt{\mu_0\rho\nu\eta}}.
\label{ReHa}
\end{equation}
{with $D=R_{\rm out}-R_{\rm in}$ and the corresponding magnetic Reynolds number $\rm Rm=Pm\ Re$ and the Lundquist number $\rm S= \sqrt{Pm}\ Ha$.}

For the rotation law with $\mu_\Om=0.25$ (the Rayleigh limit) the plots in
Fig.~\ref{f0} demonstrate the characteristic values of the solutions.
The critical Reynolds numbers (left), drift rates (middle) and wave numbers
(right) are given for the magnetic Prandtl numbers $10^{-6}$,  1 and 10, which
differ by many orders of magnitudes. The differences of the characteristic
values given in Fig. \ref{f0}, however, remain very small. In terms of the  Reynolds number  the AMRI at the
Rayleigh limit can be more easily  excited for $\rm Pm=1$ than for $\rm Pm\to0$,
but the effect is only weak compared with the variation of the magnetic Prandtl
number. Obviously, the AMRI close to the Rayleigh limit scales for very small 
\Pm\ with the Reynolds number $\rm Re$ and the Hartmann number $\rm Ha$ (Hollerbach et
al.\ 2010). As an immediate consequence, the existence of the AMRI 
can be proven by laboratory experiments with liquid metals with their 
very low values of \Pm\ -- just as the helical MRI (Stefani et al. 2009), 
but in strong contrast to the standard MRI which is much harder to achieve
(\R\ \& Hollerbach 2004). The results obtained with materials with low \Pm\ are
also representative of materials with \Pm\ of order unity. Typical values for
such possible experiments are $\rm Re\simeq 1000$ and $\rm Ha\simeq 100$ (see below).

{For $\rm Pm \gsim 1$  the mentioned scaling changes but we do not discuss these cases here in more detail. It should only be mentioned that at the Rayleigh limit the instability appears if $\Om<\Om_{\rm A}$ for very small $\rm Pm$ and  if $\Om>\Om_{\rm A}$ for very large $\rm Pm$. Both regimes are separated by the Chandrasekhar limit $\Om=\Om_{\rm A}$. The condition $\Om<\Om_{\rm A}$ for very small $\rm Pm$ can be translated into $\rm Re<Ha/\sqrt{Pm}$ which is only true for {all} $\rm Ha$ if $\rm Pm\to 0$. If, e.g., a solution exists for $\rm Ha\simeq 100$ then it is immediately clear that the corresponding $\rm Re$ for $\rm Pm=10^{-6}$ must be  {\em smaller} than $10^5$.} 

\begin{figure}
 \includegraphics[width=7cm]{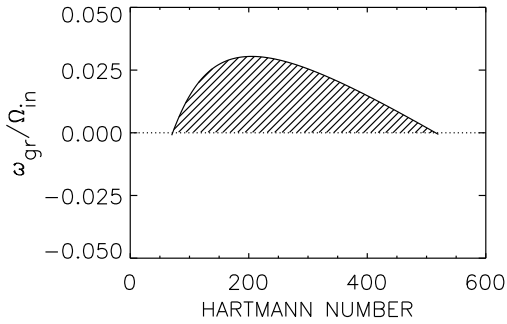}
 \caption{The (normalized) growth rates $\omega_{\rm gr}=\Im(\omega)/\Om_{\rm in}$
 as a function of the toroidal background field are only positive in the hatched domain. $m=\pm 1$.
 $\mu_\Om=0.25$, $\rm Re=3000$, ${\rm Pm}= 10^{-6}$.}
 \label{f1}
\end{figure}

The meaning of the marginal instability curves is explained by the growth
rates, given in Fig. \ref{f1} for a fixed Reynolds number $\rm Re=3000$.
The growth rate is positive between the two critical values, which in
Fig. \ref{f0} (left) enclose a region of instability. The maximum growth rate
of 0.02 lies between the two limiting Hartmann numbers, and implies a growth
time of about 7 rotation times. {Compared with the standard MRI the AMRI is slower, but exhibits the same basic scaling on the rotational timescale.}

As the left-hand plot of Fig. \ref{f0} also shows, the AMRI instability domain
is always limited by two values of the Reynolds number or two values of the
Hartmann number. For fixed rotation rate the magnetic field can be too weak or
too strong, and conversely for fixed magnetic field the rotation can be too slow
or too fast.

\begin{figure}
 \includegraphics[width=7cm]{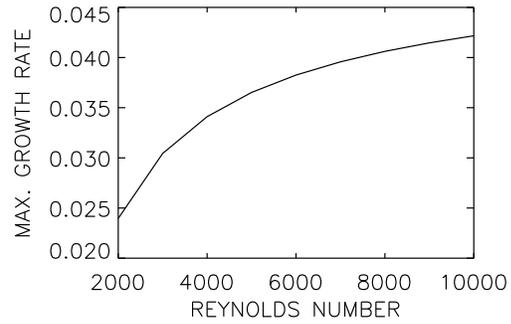}
 \caption{The {\em maximal} normalized growth rate as a function of the
Reynolds number of the basic rotation for $\mu_\Om=0.25$. The Hartmann numbers which correspond to
the maxima vary from about 200 to about 500 from the left to the right axis of
the plot. $\rm Pm=10^{-6}$.}
\label{f2}
\end{figure}

The azimuthal drift rates (Fig. \ref{f0}, middle) are always negative, hence the
pattern of the instability wave drifts in the direction of the basic rotation
(see above). A typical value of the drift (normalized with $\Om_{\rm in}$) for
small \Pm\ is 0.25 (marked in the plot), which hardly depends on the Hartmann
number. For the rotation ratio $\mu_\Om=0.25$ the
pattern essentially corotates with the outer cylinder. For stronger fields the
drift is slightly slower.

The right-hand plot in Fig. \ref{f0} gives the axial wave numbers normalized
with the gap width $D$. A wave number of $\pi$ (marked)
describes a cell with a circular geometry in the radial and the vertical
coordinates. Precisely this cell form exists for the instability at the
weak-field end of the instability domain. For stronger fields the cells become
more oblate (not prolate!) but this is only true for the lower branches of the
instability map. Along the upper branches the cells preserve their circular
shape.

One must ask how the growth rate behaves for increasingly rapid rotation. To
answer this question one must determine the maxima known from Fig. \ref{f1} for
increasing Reynolds numbers. The results given in Fig. \ref{f2} clearly show
the saturation of $\omega_{\rm gr}=\Im(\omega)/\Om_{\rm in}$. The maximum growth
rate for very rapid rotation is 0.045 $\Om_{\rm in}$, so that the shortest
growth time of AMRI at the Rayleigh limit is 3.5 rotation times.

\section{Quasi-galactic rotation}
\begin{figure}
\includegraphics[width=8cm]{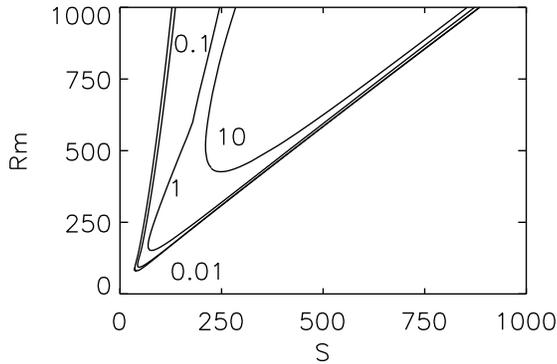}
 \caption{The instability map for the quasi-galactic rotation law $\mu_\Om=0.5$
in the $\rm S-Rm$-plane. 
The curves are marked with their values of the
magnetic Prandtl number. Curves for $\rm Pm<0.01$ are almost identical with the curve for $\rm Pm=0.01$. $m=\pm 1$.}
\label{f3}
\end{figure}

For the quasi-galactic rotation law with $\mu_\Om=0.5$, Fig.~\ref{f3} shows the
domain of instability in the ${\rm S}$-${\rm Rm}$ plane, for various magnetic
Prandtl numbers. It is obvious that for $\rm Pm\lsim 1$ the values scale with
the magnetic Reynolds number $\rm Rm$ and the Lundquist number $\rm S$.

The unstable domain in the plane again has the characteristic form of tilted
cones, so that minimal and maximal values of both ${\rm S}$ and ${\rm Rm}$ always
exist for the onset of the instability. Hence, both the magnetic field and the
rotation rate can be too weak or too strong for AMRI. One also finds that the
instability curves for ${\rm Pm}\to 0$ converge in the ${\rm S}$-${\rm Rm}$ plane. 
For all \Pm\ the ratio $\Om_{\rm A}/\Om$ lies between a low-field limit and a
high-field limit, e.g., for ${\rm Pm}=1$,
\begin{equation}
0.3\lsim \frac{\Om_{\rm A}}{\Om} \lsim 1.
\label{ratio}
\end{equation}
It is thus clear that for AMRI the Alfv\'en frequency of the toroidal magnetic
field and the rotation rate must be of the same order. If the toroidal field is
produced by differential rotation acting on a poloidal fossil field, then the
low-field limit in Eq. (\ref{ratio}) plays the role of the onset condition for
the axisymmetric instability.

\begin{figure}
 \includegraphics[width=8cm]{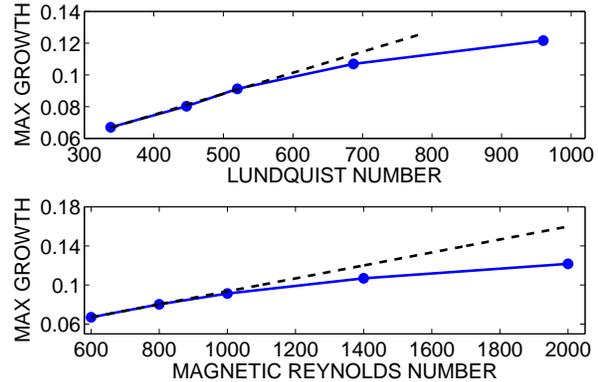}
 \caption{The  normalized growth rate $\hat\omega_{\rm gr}=\Im(\omega)/
\Om_{\rm in}$ and its saturation for $\mu_\Om=0.5$, optimized in the tilted instability cone for
$\rm Pm=1$ as a function of $\rm S$ (top) or \Rm\ (bottom).}
\label{f5}
\end{figure}

For given ${\rm Rm}$ and ${\rm Pm}$ the growth rate has been calculated for $\mu_\Om=0.5$ between
the two limiting values ${\rm S}$ where it vanishes (see also Fig. \ref{f1}); it is maximal somewhere
between the two limits. In the Fig.~\ref{f5} (bottom)  the ratio $\omega_{\rm gr}/
\Om_{\rm in}$ is plotted for various ${\rm Rm}$ and for ${\rm Pm}=1$. One finds a quasilinear relation
\begin{equation}
\frac{\omega_{\rm gr}}{\Om_{\rm in}}\simeq
 \varepsilon_{\rm gr} \, {\rm Rm} 
\label{omgr}
\end{equation}
or
\begin{equation}
\omega_{\rm gr}\simeq \varepsilon_{\rm gr}\ 
\frac{{\Om^2_{\rm in}}}{\omega_\eta} \label{omgr1}
\end{equation}
with $\varepsilon_{\rm gr}$ of order $10^{-4}$. It varies from $1.5\cdot 10^{-4}$
for ${\rm Pm}=1$ to $2.1 \cdot 10^{-4}$ for ${\rm Pm}=0.01$ (not shown), which suggests a
very weak dependence of the growth rate $\omega_{\rm gr}$ on ${\rm Pm}$. Hence,
the growth time in units of the rotation time is $\tau_{\rm gr}/\tau_{\rm rot}
\simeq 10^3/{\rm Rm}$ for small ${\rm Rm}$. For smaller ${\rm Rm}$ the AMRI is
rather slow, but the linear relation (\ref{omgr}) can only hold for small
${\rm Rm}$. One also finds  that the growth rate slowly grows for smaller
\Pm\ but this effect is rather weak.

The saturation of the normalized growth rates  is also demonstrated
by Fig. \ref{f5}. For sufficiently rapid rotation the dependence of
$\hat\omega_{\rm gr}$ on the value of \Rm\ disappears, so that
$\Im(\omega)/\Om_{\rm in}\leq 0.14$ always holds. The growth time, therefore, for the
instability of the considered rotation law and for $\rm Pm=1$ can never be
shorter than one rotation time. The same results can also be displayed in terms of the magnetic field (Fig. \ref{f5}, top).
\section{An experiment}
\begin{figure}
 \includegraphics[width=8cm]{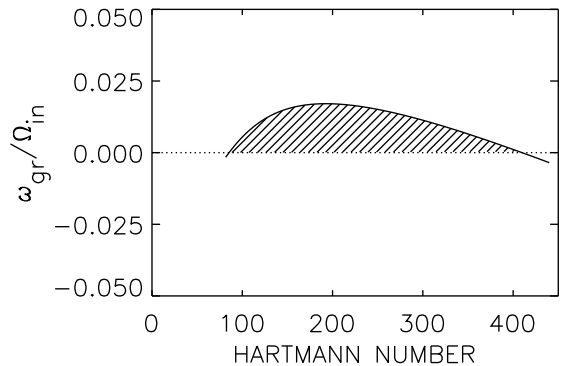}
 \caption{The normalized growth rate for an experiment with $\mu_\Om=0.26$ as a
function of the toroidal background field. $m=\pm 1$, $\rm Re=3000$, ${\rm Pm}=
10^{-6}$. They are positive only in the hatched domain. 
There are no visible differences to the results for $\rm Pm=0$.}
\label{f6}
\end{figure}

With the given data it is easy to design an experiment for the realization of
AMRI in the laboratory. For experiments on magnetic-induced instabilities of
the differential rotation it is natural to work with a
Taylor-Couette flow between two rotating cylinders and with a rotation law which by itself
is hydrodynamically stable. On the other hand, it is reasonable to apply a rotation law very close to the Rayleigh limit because for small \Pm\ the critical rotation rate scales (only) with $\rm Re$. Figure \ref{f6} shows that for $\mu_\Om=0.26$ the
limit of marginal instability is reached for  a Hartmann number of 85 (and  $\rm Re=3000$). Note that
this instability region is slightly smaller than for the rotation law with
$\mu_\Om=0.25$ (which is understandable as the energy for AMRI is completely
provided by the energy of the shear). 

The relation $I_{\rm axis}= 5 R_{\rm in} B_{\rm in}$ connects the toroidal field
amplitude $B_{\rm in}$ at $R_{\rm in}$ with the axial current inside the inner
cylinder. $I$, $R$ and $B$ must be measured in ampere, cm and gauss. Hence,
\begin{equation}
{\rm Ha}= \frac{1}{5}\frac{I_{\rm axis}}{\sqrt{\mu_0\rho\nu\eta}}.
\label{Ha}
\end{equation}
The radial size of the container does not occur in this relation between the
Hartmann number and the current amplitude. For the gallium alloy GaInSn the
magnetic Prandtl number is $1.4\cdot 10^{-6}$ and the value of the square root
in (\ref{Ha}) is 25.6. The resulting electric current for marginal instability
is thus 10.9 kA.

\begin{figure}
\hskip-1cm
\hbox{ 
\includegraphics[width=4.8cm]{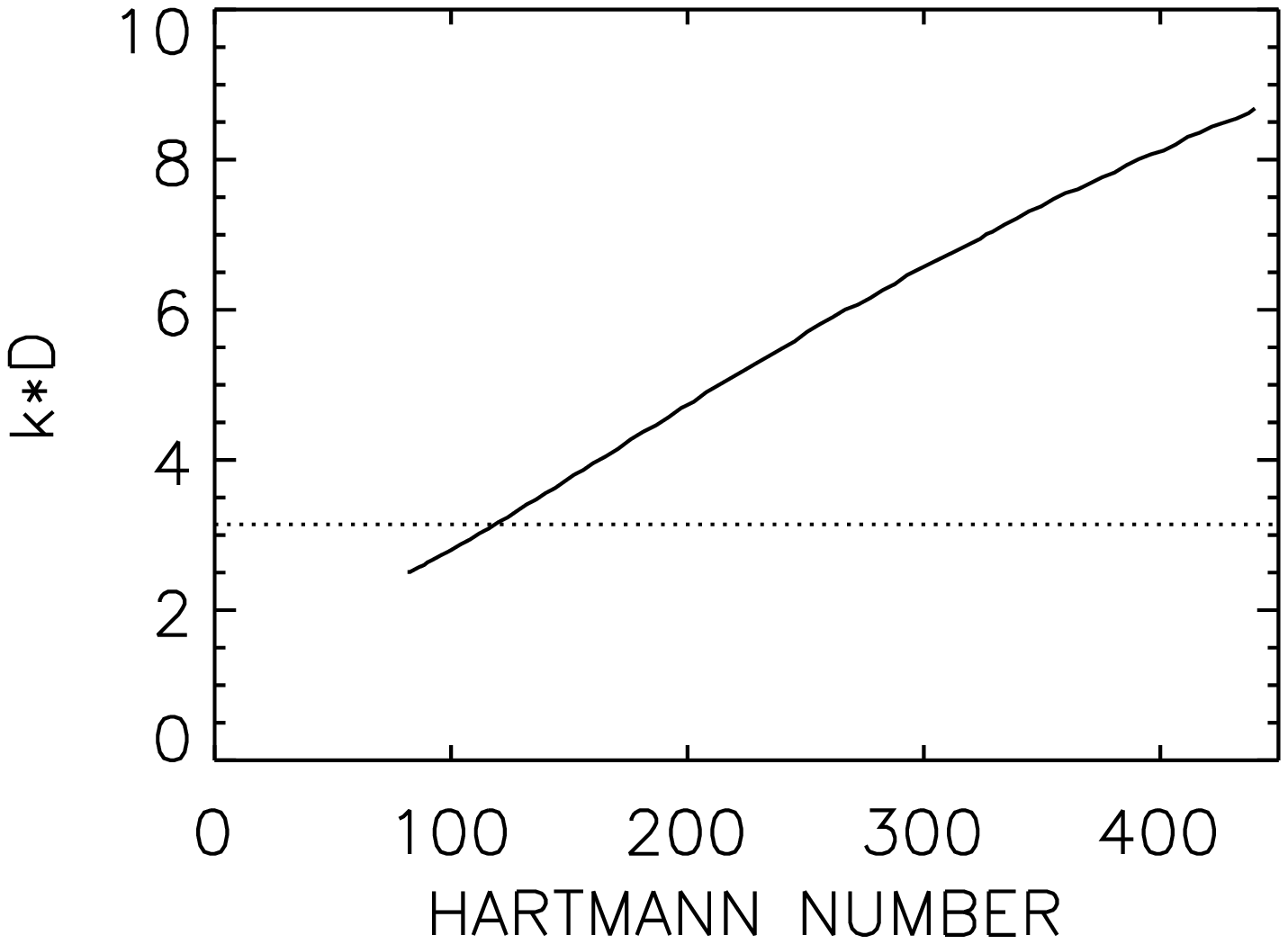}
\includegraphics[width=4.8cm]{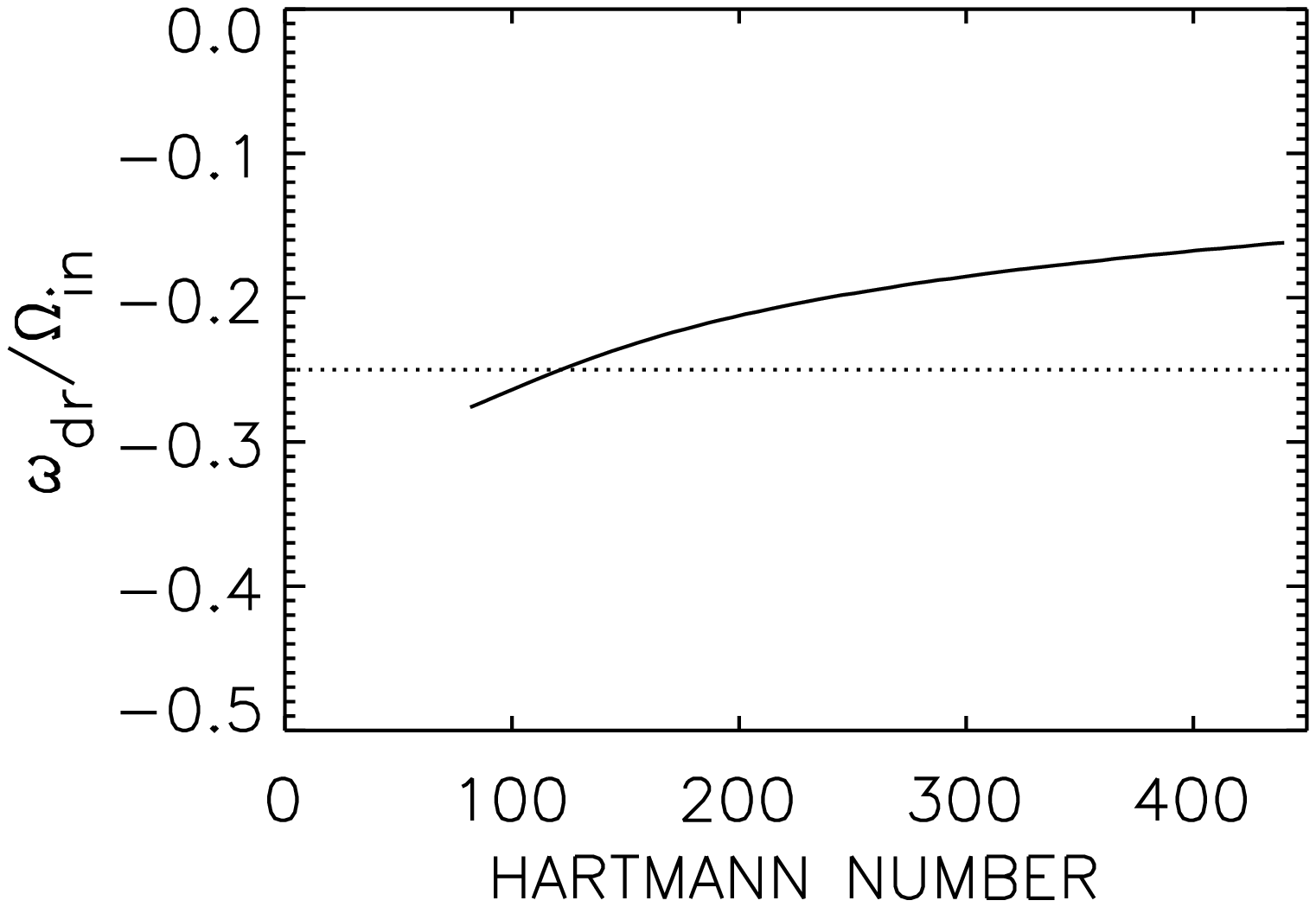}
}
 \caption{The normalized wave number (left) and drift rate (right) for fixed
Reynolds number $\rm Re=3000$. The dotted lines give the wave number ($\pi$)
for a circular cell structure (left) and the drift rate of $-0.25$ for
corotation of the pattern with the outer cylinder (right). $m=\pm 1$,
$\mu_\Om=0.26$, ${\rm Pm}= 10^{-6}$.}
\label{fX}
\end{figure}

For a container with (say) $R_{\rm in}=4$\ cm, $R_{\rm out}=8$\ cm and with the viscosity of GaInSn
($\nu=3.4\cdot 10^{-3}$\ cm$^2$/s) the Reynolds number $\rm Re=3000$ requires a
rotation rate of the inner cylinder of only 0.1 Hz. Figure \ref{fX} shows that
for the weakest fields the cell structure of the nonaxisymmetric pattern is
circular and that the pattern drifts with nearly the same rotation rate as the
outer cylinder. Both properties are very characteristic for the instability
close to the marginal limit for weak magnetic fields -- independent of the
steepness $\mu_\Om$ of the rotation law.
Such an experiment is currently operating at the Helmholtz-Zentrum
Dresden-Rossendorf. The results will be presented in a separate paper.

\section{Kinetic and magnetic energies}
\begin{figure}
 \includegraphics[width=6cm]{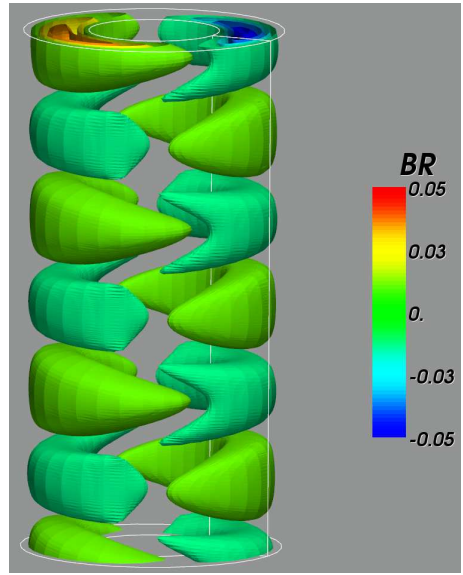}
 \caption{The nonlinear instability pattern of AMRI is a combination of the modes
$m=\pm 1$ which drifts with a common drift speed ${\rm d}\phi/{\rm d}t$. There are seven cells along the
axial direction with the aspect ratio $\Gamma=8$. The cells are thus slightly
elongated. $\rm Re=500$, $\rm Ha=80$, $\mu_\Om=0.26$, ${\rm Pm}=0.1$. }
\label{f7}
\end{figure}

We have seen that the modes for $m=1$ and $m=-1$ (corresponding to left and right
spiraling modes, Hollerbach et al. 2010) are fully degenerate, i.e. they are
excited at the same eigenvalues and possess the same wave numbers and azimuthal phase speeds. To find the resulting instability pattern nonlinear calculations are necessary. 
 For the following  calculations the spectral cylindrical MHD code
of Gellert, R\"udiger \& Fournier (2007) is used. The solution is expanded in
azimuthal Fourier modes. The resulting meridional problems are solved with a
Legendre spectral element method after Deville, Fischer \& Mund (2002). The code,
however, is only able to solve equations with a minimum magnetic Prandtl number
of 0.01. Fig. \ref{f7} shows the structure of the resulting wave, which consists
of both $m=\pm1$ components, and drifts together with the outer cylinder. There are seven cells along the vertical axis with its eight length units.  The
almost circular meridional cell structure thus  confirms the results of the linear
theory.

Besides the geometric structure, the nonlinear code also provides the behavior of
the equilibrated energies. The main question is  how strong the magnetic energy is,
in units of the kinetic energy. Note that  the maximal amplitude of the radial field component in units of
$B_{\rm in}$ in Fig. \ref{f7} is rather small. This fact, however, is only due to
the rather slow rotation of this example. 
\begin{figure}
 \includegraphics[width=8cm]{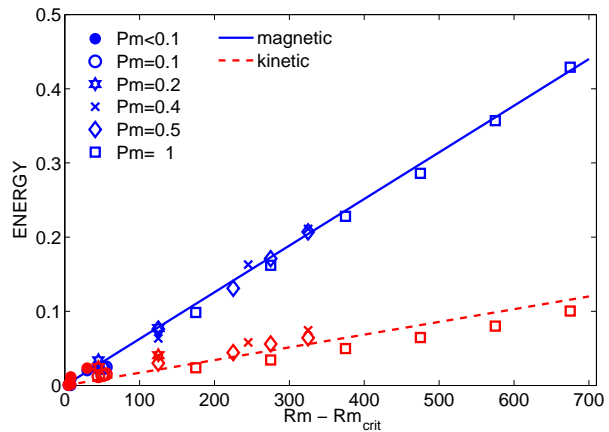}
 \caption{The magnetic (solid) and the kinetic (dashed) energies of AMRI for
various magnetic Prandtl numbers vs. the magnetic Reynolds numbers.
$\rm Re=1000$, $\rm Ha=75\dots 200$, $\mu_\Om=0.26$, ${\rm Pm}= 0.05\dots 1$. }
\label{f8}
\end{figure}

Figure \ref{f8} displays the energies of several instability realizations for
magnetic Prandtl numbers between 0.05 and 1.0. Both the magnetic and the
kinetic energy are  normalized with the energy of the toroidal
background field. All energies are integrals over the entire container. 
The first finding concerns the quantity ${\rm q}=
\langle \vec{b}^2\rangle/ B_{\rm in}^2$,
which for driven turbulence in the high-conductivity limit  scales with
$\rm Rm$ (Br\"auer \& Krause 1974). The same is true for this quantity in the
case of AMRI (Fig. \ref{f8}). The  relation  
\begin{equation}
{\rm q}= \varepsilon_{\rm mag}\ ({\rm Rm-Rm_{crit}})
\label{q}
\end{equation}
has been found with $\varepsilon_{\rm mag}\simeq 6\cdot 10^{-4}$. It should thus be possible
that the energy of the magnetic fluctuations reaches the order of the magnetic
background field. The equation (\ref{q}) can also be written as ${\rm q}\propto
\varepsilon_{\rm mag} \Om/\omega_\eta$. The saturation of $\rm q$
 for large magnetic Reynolds numbers, however, is not yet known.
A possible form of the relation for fast rotation might be ${\rm q}\propto \Om^2_{\rm A}/\Om^2$.

Figure \ref{f8} also shows that the magnetic energy always dominates the kinetic
energy of the fluctuations. One may speculate that the factor of four which
arises here also appears in the ratio of the resulting eddy viscosity and the
chemical diffusivity (see below). 
\begin{figure}
 \includegraphics[width=8cm]{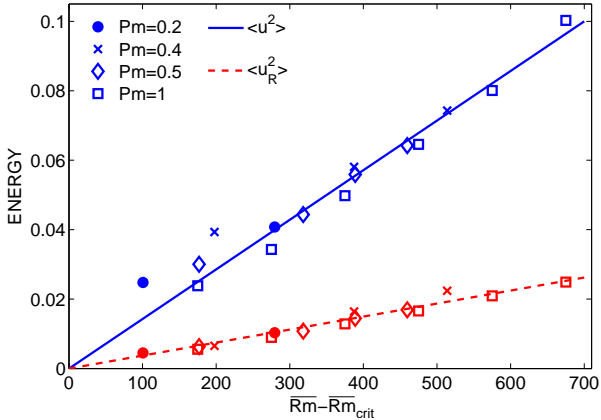}
 \caption{The kinetic energy (solid) and its portion in the radial direction
(dashed) -- both normalized with the energy of the magnetic background field --  of the AMRI perturbations for various \Pm\ vs. the modified magnetic
Reynolds number ${\overline {\rm Rm}}$. Note that the radial intensity is almost 1/3 of the total intensity $\langle \vec{u}^2\rangle$.
 $\rm Ha=75\dots 200$, $\mu_\Om=0.26$.}
 \label{f9}
\end{figure}
According to Fig. \ref{f9} the kinetic energy for various magnetic Prandtl
numbers vs. the modified Reynolds 
number ${\overline {\rm Rm}}
=\Om_{\rm in}R_{\rm in}^2/\bar\eta$ with $\bar\eta=\sqrt{\nu\eta}$ provides the
linear relation
\begin{equation}
\frac{\mu_0\rho\langle \vec{u}^2\rangle}{B^2_{\rm in}}=
 \varepsilon_{\rm kin}\ ({\overline {\rm Rm}} -{\overline {\rm Rm}_{\rm crit}}) 
\label{kin1}
\end{equation}
with $\varepsilon_{\rm kin}\simeq 1.4\cdot 10^{-4}$. Hence,
 \begin{equation}
\frac{u_{\rm rms}}{\Om_{\rm in}R_{\rm in}}\simeq 0.012\ {\rm {\bar \Lambda}}^{0.5}
\label{kin2}
\end{equation}
with the modified Elsasser number 
 \begin{equation}
 {\rm {\bar\Lambda}}= \frac{B^2_{\rm in}}{\mu_0\rho\Om\bar\eta},
\label{kin3}
\end{equation}
which again does {\em not} depend on the size of the container. The appearance of the
molecular viscosity in the expression of the kinetic energy might have strong
consequences for the value of the kinetic energy.  For not too small \Pm, however, it is clear that the AMRI
provides higher values of the magnetic energy compared with the kinetic energy. This means that the 
resulting eddy viscosity also exceeds  the mixing coefficient of chemicals.
The angular momentum transport is dominated by the magnetic energy while
magnetic fluctuations do not contribute to the diffusion of thermal energy
and/or the mixing of chemicals (Vainshtein \& Kichatinov 1983).
\begin{figure}
 \includegraphics[width=8cm]{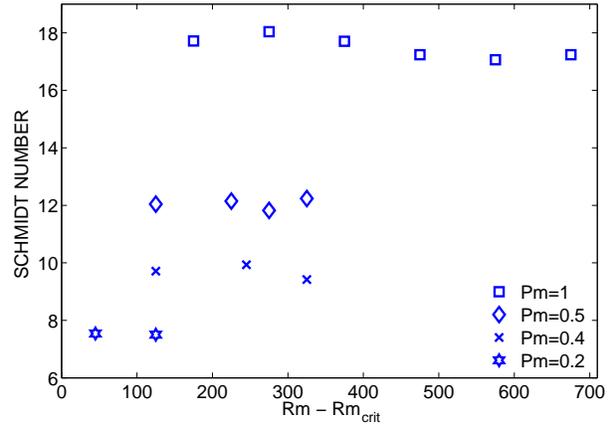}
 \caption{The proxy (\ref{epsil}) of the turbulent Schmidt number for various
magnetic Prandtl numbers and for $\mu_\Om=0.26$ does {\em not} depend on the magnetic
Reynolds number but decreases as $\sqrt{\rm Pm}$.}
 \label{f10}
\end{figure}

The different scaling of the magnetic and kinetic energies has consequences for
their ratio $\epsilon=\langle \vec{b}^2\rangle/\mu_0\rho\langle \vec{u^2}\rangle$,
which after (\ref{q}) and (\ref{kin1}) scales as $\sqrt{\rm Pm}$. The magnetic energy only dominates for
magnetic Prandtl numbers of order unity or larger. Consequently, the turbulent
Schmidt number 
\begin{equation}
{\rm Sc}=\frac{\nu_{\rm T}}{D_{\rm T}}\simeq \epsilon_R
\label{sc}
\end{equation}
should also scale with $\sqrt{\rm Pm}$. Here $\epsilon_R$ is formed with the
radial velocity components (also given in Fig. \ref{f9}), i.e. 
\begin{equation}
\epsilon_R=\frac{\langle \vec{b}^2\rangle}{\mu_0\rho\langle {u_R^2}\rangle},
\label{epsil}
\end{equation} 
as only the radial turbulent intensity is responsible for the radial mixing of
chemicals. Figure \ref{f10} demonstrates that indeed the turbulent Schmidt number
on the basis of AMRI strongly exceeds unity, but decreases with $\sqrt{\rm Pm}$
for small $\rm Pm$. There is no clear evidence for a strong dependence of the Schmidt number on the basic rotation. 
Note, however, that for very small magnetic Prandtl number it
cannot become smaller than 0.4 (Yousef et al. 2003). In fluids  with small magnetic
Prandtl number the angular momentum transport and the mixing of chemicals by AMRI
are of the same power. Both processes   stop when  rigid
rotation is reached. Only as long as the differential rotation is strong  the
mixing of chemicals is also strong. It might be, however, that the inclusion of
the `negative' buoyancy by the stable stratification of real stellar cores only
reduces the mixing rather than the angular momentum transport and the Schmidt
number is not reduced too much by the small magnetic Prandtl number (Kitchatinov
\& Brandenburg 2012).

\section*{Acknowledgments}  This work was supported by the Deutsche
Forschungsgemeinschaft (SPP 1488 {\em PlanetMag}).  RH is supported at ETH Z\"urich by ERC Grant No. 247303 (MFECE).  

 

\bsp

\label{lastpage}

\end{document}